\documentclass[journal,10pt,letterpaper,final,twocolumn]{IEEEtran}%
\usepackage{amsmath}
\usepackage{amsfonts}
\usepackage{cite}
\usepackage{amssymb}
\usepackage{mathrsfs}
\usepackage{graphicx}
\usepackage[usenames,dvipsnames]{color}
\usepackage{epstopdf}
\usepackage[all]{xy}
\usepackage[]{mcode}
\usepackage{supertabular}
\usepackage{subcaption}
\usepackage{array}

\providecommand{\U}[1]{\protect\rule{.1in}{.1in}}
\pdfoutput=1
\providecommand{\U}[1]{\protect\rule{.1in}{.1in}}

\newcommand{\qed}{\nobreak \ifvmode \relax \else
      \ifdim\lastskip<1.5em \hskip-\lastskip
      \hskip1.5em plus0em minus0.5em \fi \nobreak
      \vrule height0.75em width0.5em depth0.25em\fi}

\begin{document}

\title{\LARGE Toward Spectral and Energy Efficient 5G Networks Using Relayed OFDM with Index Modulation}
\author{Shuping Dang, \textit{Member, IEEE}, Basem Shihada, \textit{Senior Member, IEEE}, Mohamed-Slim Alouini, \textit{Fellow, IEEE}
  \thanks{The authors are with Computer, Electrical and Mathematical Science and Engineering Division, King Abdullah University of Science and Technology (KAUST), 
Thuwal 23955-6900, Kingdom of Saudi Arabia (e-mail: \{shuping.dang, basem.shihada, slim.alouini\}@kaust.edu.sa).}}

\maketitle

\begin{abstract}
Next generation wireless networks are expected to provide much higher data throughput and reliable connections for a far larger number of wireless service subscribers and machine-type nodes, which result in increasingly stringent requirements of spectral efficiency (SE) and energy efficiency (EE). Orthogonal frequency-division multiplexing with index modulation (OFDM-IM) stands out as a promising solution to satisfy the SE requirement with a reasonable increase in system complexity. However, the EE of OFDM-IM is still required to be enhanced. Moreover, diversity gain is hardly harvested in OFDM-IM systems, which hinders further reliability enhancement. In this regard, relay assisted OFDM-IM, as a promising joint paradigm to achieve both high SE and EE, was proposed and has been studied since last year. The objectives of this article is to summarize the recent achievements of this joint paradigm, articulate its pros and cons, and reveal the corresponding challenges and future work.
\end{abstract}

\section*{Introduction}
Current study has foreboded the tendency that data throughput and the number of connected nodes in next generation wireless networks will tremendously increase, which results in increasingly stringent requirements of spectral efficiency (SE) and energy efficiency (EE). To meet these two requirements, orthogonal frequency-division multiplexing with index modulation (OFDM-IM) attracts researchers' attention in recent years since being proposed in \cite{6587554}. Different from conventional amplitude-phase modulation (APM) schemes, OFDM-IM employs an \textit{index domain} in addition to the classic amplitude-phase constellation diagram, so as to form a three-dimensional modulation scheme, which considerably enhances the SE under proper system configurations \cite{8004416}.

By OFDM-IM, only a subset of orthogonal subcarriers will be activated to form a unique subcarrier activation pattern (SAP), which can be generated by subcarrier grouping and inverse fast Fourier transform (IFFT) \cite{6841601}. Consequently, we can make the number of legitimate SAPs a power of two, and resort to the index of SAP to modulate extra bit stream in addition to the bit stream modulated by data constellation symbols carried on active subcarriers. Recent studies have also proved that under appropriate system configurations, OFDM-IM is superior to plain OFDM in terms of SE and/or error performance \cite{7469311}.

Although OFDM-IM can gain a higher SE, its EE is still required to be enhanced. In addition, diversity gain is hardly harvested in OFDM-IM systems, which hinders further reliability enhancement for OFDM-IM systems. To manage these rising issues with OFDM-IM and achieve high SE and EE simultaneously, cooperative multi-hop architecture consisting of relay(s) is incorporated with OFDM-IM systems in \cite{8075970,8358694}, and the resultant relay assisted OFDM-IM has sparked the research interest in academia within a short period of time. A number of papers have shown that the reliability of OFDM-IM can be improved by incorporating the cooperative multi-hop architecture, and a higher degree of system flexibility is obtainable. In particular, the optimization and design dimensions can be extended from the frequency domain to the spatial domain, which facilitates more advanced techniques to be employed, for example cognitive radio (CR), subcarrier permutation (SP), power allocation (PA), relay selection (RS), and full-duplex (FD) relaying \cite{8227748,8241721,8269167,8476574,8405601,8361430,8614439,8612925,8661773}. 

In this regard, the objectives of this article is to summarize the recent achievements of relay assisted OFDM-IM, articulate pros and cons of this new paradigm, and reveal the corresponding challenges and future work for further research activities. In this article, we only focus on a single group of $N$ subcarriers and adopt the simplistic look-up table method with a fixed number of $K$ active subcarriers for bit-to-SAP mapping. The APM scheme is assumed to be $M$-ary phase-shift keying (PSK). The numbers of hops and relays in a single hop are denoted as $L$ and $T$, respectively. 

\begin{figure*}[!t]
\centering
\includegraphics[width=5.5in]{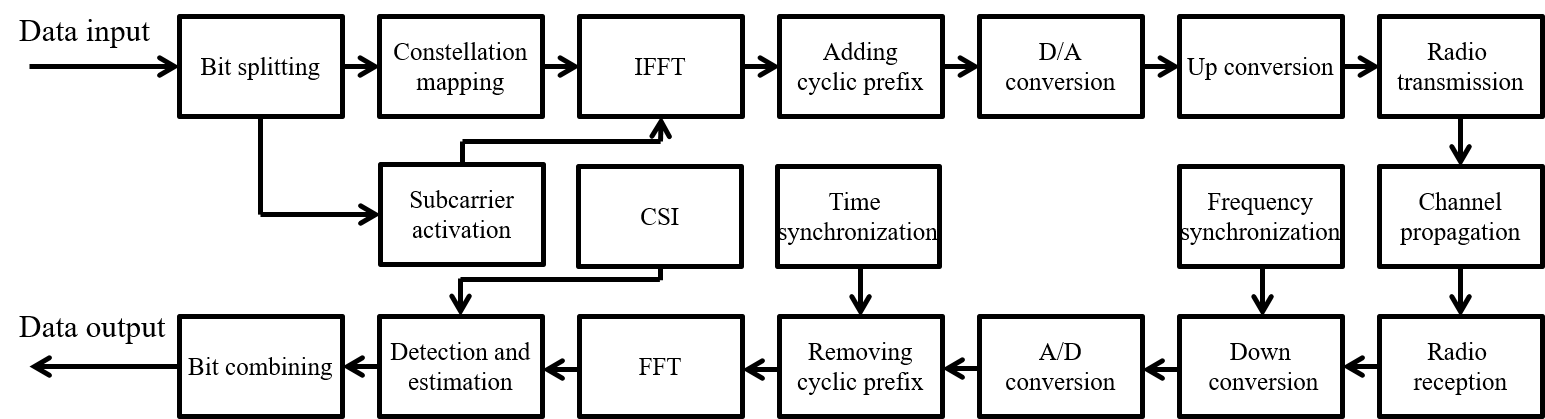}
\caption{System diagram of a typical DF relay assisted OFDM-IM system.}
\label{sys}
\end{figure*}

\section*{Overview and Application Scenarios}
The basic principle of relay assisted OFDM-IM is similar to classic/plain OFDM supported by cooperative multi-hop architecture. However, only a subset of subcarriers will be received and forwarded by relay(s) to the destination. A complete system diagram of a typical decode-and-forward (DF) relay assisted OFDM-IM system is pictorially presented in Fig. \ref{sys}. For amplify-and-forward (AF) relay assisted OFDM-IM, the receiving part from `Down conversion' to `Data output' modules can be simply replaced by a signal amplifier that could also be supported by channel state information (CSI) depending on whether fixed-gain AF or variable-gain AF relaying protocol is applied.

\begin{figure}[!t]
\centering
\includegraphics[width=3.2in]{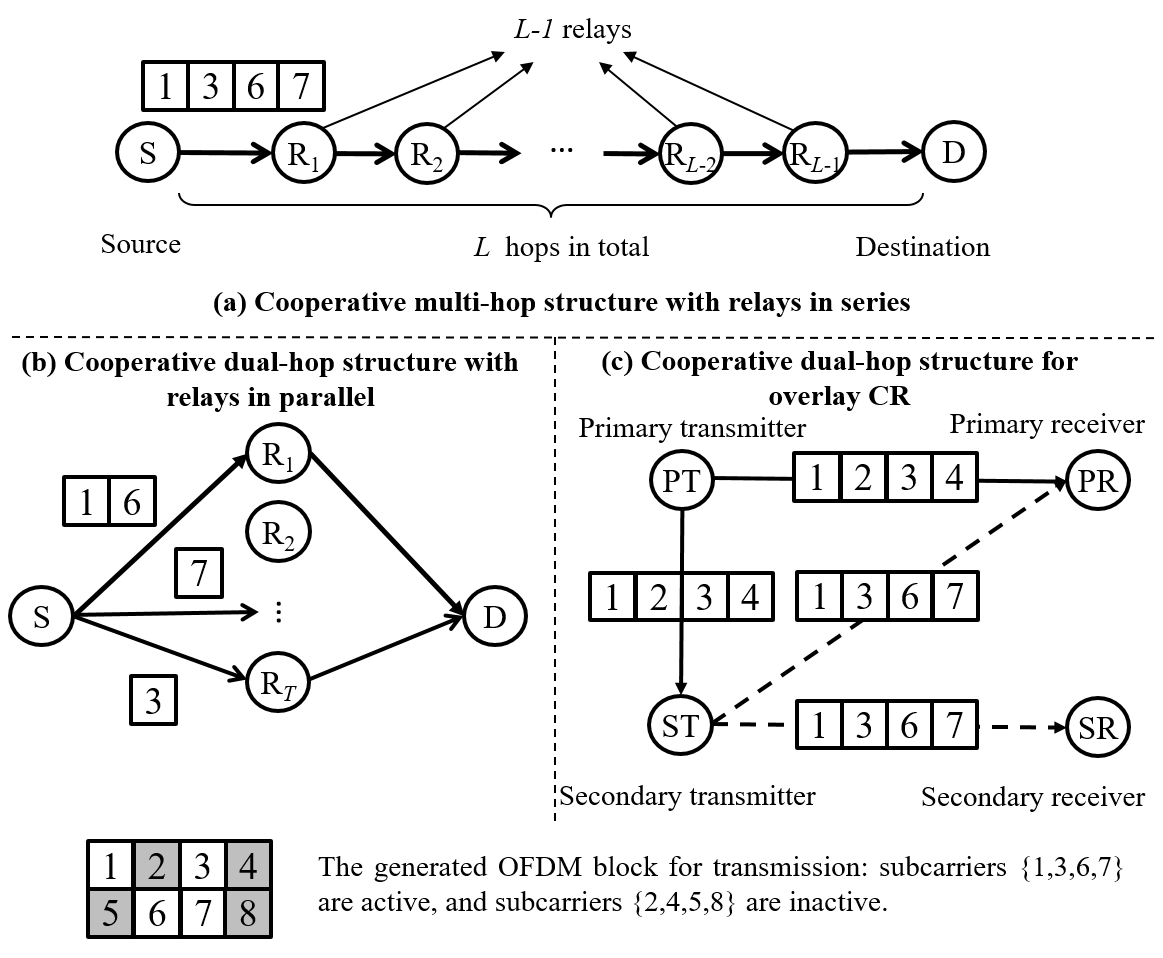}
\caption{Three major cooperative multi-hop structures for relay assisted OFDM-IM.}
\label{structure}
\end{figure}

Depending on the design of cooperative multi-hop structure, we can have a number of system models for applying relay assisted OFDM. There are three main cooperative multi-hop structures that are practical for relay assisted OFDM-IM systems: 
\begin{itemize}
\item Cooperative multi-hop structure with relays in series
\item Cooperative dual-hop structure with relays in parallel
\item Cooperative dual-hop structure for overlay CR
\end{itemize}
Most of other structures for relay assisted OFDM-IM can be regarded as special cases or combinations of these three major structures. For clarity, we illustrate them in Fig. \ref{structure}, and expatiate on their features  as follows.

\subsection*{Cooperative Multi-Hop Structure with Relays in Series}
The cooperative multi-hop structure with $L-1$ relays in series (i.e., $L$ hops) is suited for enabling long-distance transmission of OFDM block, and assumes transmission links only exist between two adjacent nodes. This structure is investigated in \cite{8075970,8241721,8358694,8269167,8476574} and is the most common structure for relay assisted OFDM-IM. By such a multi-hop structure, ideally, the network coverage can be extended to arbitrarily large as long as a sufficient number of relays are deployed in a proper manner. As a consequence, relay assisted OFDM-IM supported by the cooperative multi-hop structure with relays in series would have a higher SE and EE. On the other hand, if DF relaying protocol is adopted over this structure, error propagation issue should be taken into consideration, since the correct detection at the destination is based on the prerequisite that all $L-1$ relays must correctly detect the received OFDM block, which becomes less likely when the number of hops is large. On the contrary, if AF relaying protocol is adopted, noise amplification and nonlinear distortion caused by amplification saturation would be severe, especially for large $L$.

\subsection*{Cooperative Dual-Hop Structure with Relays in Parallel}
To mitigate the disadvantages of cooperative multi-hop structure with relays in series and provide a higher degree of network design flexibility, the cooperative dual-hop structure with $T$ relays in parallel is in use, when the distance between source and destination is moderate \cite{8405601,8361430,8614439,8612925}. Another benefit of the cooperative dual-hop structure with relays in parallel is that a variety of RS schemes can be performed to harvest a coding gain and/or even a diversity gain. On the contrary, the rising issue of this structure is the increasing system complexity and signaling overhead caused by multi-relay coordination and synchronization. First, in order to select appropriate relay(s), full or partial CSI must be known by the source node in a certain way. This can be achieved by a centralized manner via a central node (e.g., base station (BS) and access point (AP)) or a timer-based distributed manner that waits for the response from relays. The former would require an extra feedback channel exclusively used for selected relay indication and latter could cause an additional delay. The synchronization issue comes from the fact that multiple relays are spatially distributed and operate in different oscillators.

\subsection*{Cooperative Dual-Hop Structure for Overlay CR}
The cooperative dual-hop structure for overlay CR is proposed in \cite{8227748}, which considers both primary and secondary user pairs. Specifically, in the first phase, the primary transmitter (PT) transmits its information to both primary receiver (PR) and the secondary transmitter (ST). The ST acting as a relay for the primary transmission re-transmits the signal from the PT to the PR. However, different from conventional CR cooperative networks, OFDM-IM is utilized at the ST, and the ST split the transmission space into the classic phase-amplitude constellation domain and the index domain. The information required to be re-transmitted is modulated in the classic phase-amplitude constellation domain, while the information intended to be transmitted to secondary receiver (SR) is encoded in the index domain via the SAP. Therefore, the OFDM block transmitted by the ST in the second phase can both help the decoding process in the PR for the primary transmission, and concurrently piggyback information for the secondary transmission to the SR. The mutual interference between primary and secondary transmissions can thus be alleviated.

\section*{State-of-the-Art Achievements}
It has been pointed out in \cite{8004416} that due to the low implementation cost and improved SE as well as extended coverage, cooperative communications would be a helpful ally of OFDM-IM, and the combination of both prototypes is worth investigating. The initial work introducing the cooperative dual-hop architecture with a single DF relay to OFDM-IM systems is published in \cite{8075970}. Only numerical results generated by Monte Carlo simulations are shown in this work to testify the performance superiority of this combination. From these preliminary results, several key properties of relay assisted OFDM-IM can be observed and it is thus summarized that relay assisted OFDM-IM would be considered as an intriguing solution for both long-range energy-efficient communications and high data rate communications on  cell edges.

In the meantime, OFDM-IM-aided cooperative relaying protocol for CR networks is proposed in \cite{8227748}, in which the cooperative dual-hop structure for overlay CR is in use. Upper bounds on bit error rate (BER) of both primary and secondary transmissions are derived in single-integral form and the achievable rates of the proposed system with a variety of finite-size constellations are also analyzed. Numerical results verify the analysis presented and meanwhile confirm that the error performance of both primary and secondary user pairs is improved by adopting OFDM-IM over the cooperative dual-hop structure for overlay CR.

Following these pioneering papers, dual-hop relay assisted OFDM-IM with SP is analyzed in \cite{8241721}. In particular, the codebook mapping incoming bits to SAPs is dynamically changed according to CSI, and it is allowed to perform SP at the relay node. In this way, the end-to-end performance consisting of two hops can be decoupled and a higher degree of freedom is achievable, which leads to an enhanced reliability and a frequency diversity gain. The analysis pertaining to outage performance, error performance and network capacity is provided and corroborated by numerical results. PA is further incorporated in the dual-hop relay assisted OFDM-IM with SP in \cite{8269167}, and the formulated power allocation problem over active subcarriers can be solved based on the Karush-Kuhn-Tucker (KKT) conditions.

The cooperative dual-hop model is generalized to cooperative multi-hop model in \cite{8358694}. Because of the employment of DF relaying, the bottleneck effect will yield a constraint on the number of hops, as with an increasing number of hops, the outage and error performance will be considerably degraded. An ingenious analytical methodology based on the concept of end-to-end link is used so as to facilitate the performance analysis of relay assisted OFDM-IM in cooperative multi-hop networks. In addition, as a minor contribution, the transmission rate of relay assisted OFDM-IM is analyzed in depth. To mitigate the bottleneck effect, AF relaying protocol is integrated into OFDM-IM and analyzed thoroughly in \cite{8661773}.

Multi-relay assisted OFDM-IM enhanced by RS are investigated in \cite{8405601,8361430,8614439,8612925}. They adopt the cooperative dual-hop structure with relays in parallel and employ multi-carrier RS schemes. \cite{8405601} assumes that only the CSI in the first hop is accessible at the source for RS purposes, so that a partial RS (PRS) scheme based on centralized control is formed. To be realistic, the channel estimation error is taken into consideration in this paper, which yields imperfect CSI for PRS. \cite{8361430,8614439,8612925} suppose that full CSI is perfectly known and thereby the end-to-end channel power gain can be adopted as the indicator for RS. As a result, two common multi-carrier RS schemes for classic OFDM systems, bulk and per-subcarrier (PS) RS schemes are applied. \cite{8612925} in particular involves the AF relaying protocol with multi-relay selection in relay assisted OFDM-IM. All aforementioned papers have shown that spatial coding and/or diversity gains for relay assisted OFDM-IM can be harvested by RS.

All aforementioned papers adopt half-duplex DF relaying as the forwarding protocol at relay(s). Apart from half-duplex relaying, full-duplex DF relay assisted OFDM-IM is proposed and analyzed in \cite{8476574}, in which residual self-interference (RSI) at the full-duplex relay node is investigated and its power is modeled as an exponentially distributed random variable. Most different from half-duplex relay assisted OFDM-IM, the transmission in full-duplex relay assisted OFDM-IM will be affected by previous transmission attempts, which cause a difference in RSI over frequency bands because different SAPs are employed in different transmission attempts.

\section*{Advantages and Disadvantages}
From an engineering perspective, any technology/paradigm can gain in one aspect at the cost of the other. The same rule applies to relay assisted OFDM-IM. As we have more or less mentioned earlier, we elaborate the main advantages and disadvantages of relay assisted OFDM-IM infra, so as to provide a full picture of this new paradigm. Subsequently, simulation results for a case study are demonstrated to provide quantitative information and an insight into various relay assisted OFDM-IM systems.

\subsection*{Advantages}
\subsubsection*{Diversity and Coding Gains} 
Without involving channel coding techniques, the diversity gain of relay assisted OFDM-IM systems can be harvested from the frequency domain and/or the spatial domain by SP and RS, respectively. Both diversity and coding gains can enhance OFDM-IM systems in terms of outage and error performance.

\subsubsection*{Energy Efficiency}
Since the radio wave propagation distance can be effectively shrunken from a complete one to $L$ segments by cooperative relaying, the required transmit power to achieve a given quality of service (QoS) is much smaller than that without support of relays. As a consequence, the EE superiority of relay assisted OFDM-IM can be easily deduced from the Friis transmission equation.

\subsubsection*{Spectral Efficiency}
The SE of OFDM-IM systems can also be further enhanced as a concomitant of EE enhancement by introducing cooperative relaying. This is simply because the reduction of transmit power will allow the reuse of the same spectral resources among multiple users in proximity. From a holistic viewpoint, the network throughput is enlarged without utilizing extra spectral resource and thereby the SE rises.

\subsection*{Disadvantages}
\subsubsection*{Transmission Delay}
To decouple the transmission over $L$ hops and hence prevent the cross-hop interference, $L$ orthogonal time slots for one complete transmission from source to destination are required, for half-duplex relaying. This would result in a considerable transmission delay. Also, if DF relaying is adopted as the forwarding protocol at relays, an additional delay is yielded by the decoding and re-encoding process.

\subsubsection*{Complexity and Signaling Overhead}
The implementations of cooperative multi-hop architecture and multiple relays inevitably render a higher system complexity and signaling overhead. First, time and frequency synchronizations are required among multiple nodes and the multi-relay coordination on a distributed basis is demanding. Second, channel estimation in cooperative multi-hop  networks is also difficult and causes extra signaling overhead, especially when channels are volatile. Depending on the network protocol and application scenario, multiple replicas are received at the destination and shall be combined before processing, which again rises the system complexity.

\subsubsection*{Error Propagation}
Error propagation is a unique issue for DF relay assisted OFDM-IM only, indicating the scenario that an erroneously decoded OFDM block by an intermediate DF relay is re-transmitted and detected by the posterior relay(s) and the final destination. This also refers to the bottleneck effect of DF relaying and should be dealt with carefully.

\subsubsection*{Nonlinear Distortion}
Contrary to error propagation for DF relay assisted OFDM-IM, nonlinear distortion is an exclusive arduousness for AF relay assisted OFDM-IM. Nonlinear distortion is caused by the amplification saturation to the nonlinear region. Because most electronic components can only operate linearly within a limited power region, this phenomenon might yield severe inter-channel interference (ICI) and a deleterious impact on signal detection.

For clarify, we provide qualitative comparisons of a number of crucial properties among plain OFDM, classic OFDM-IM and relay-assisted OFDM-IM in Table \ref{comparison}.

\begin{table}[!t]
\renewcommand{\arraystretch}{1.3}
\caption{Qualitative Comparisons among plain OFDM, classic OFDM-IM and relay-assisted OFDM-IM.}
\label{comparison}
\centering
\begin{tabular}{c|c|c|c}
\textbf{Key measures} & \begin{tabular}{@{}c@{}}Plain \\ OFDM\end{tabular} & \begin{tabular}{@{}c@{}}Classic \\ OFDM-IM\end{tabular} & \begin{tabular}{@{}c@{}}Relay assisted \\ OFDM-IM\end{tabular}\\
\hline\hline
Reliability & Moderate & Moderate & High \\
Throughput & Low & Moderate & Moderate \\
Spectral efficiency  & Moderate & Moderate & High \\
Energy efficiency & Low & Moderate & High \\
System complexity & Low & Moderate & High \\
Signaling overhead & Low & Low & Moderate \\
Transmission delay &  Low & Low & High \\
\hline

\hline
\end{tabular}
\end{table}

\subsection*{Case Study}
To pictorially illustrate the performance superiority of relay assisted OFDM-IM and reveal the effects of different network structures and system configurations, we carried out a series of simulations regarding block error rate (BLER), BER, outage probability (OP), and throughput in bit per channel use (bpcu) for various cases. To simplify the simulations and restrict our discussions within a reasonable scope in this article, we make the following assumptions when carrying out simulations:
\begin{itemize}
\item A sufficiently long CP and perfect synchronizations in time and frequency are implemented, so as to eliminate inter-symbol interference (ISI) and ICI.
\item Instantaneous CSI is perfectly accessible to all nodes to facilitate RS (if required) and estimation. 
\item Free space is supposed to be the signal propagation environment and the path loss exponent is thereby $\alpha=2$.
\item All nodes are physically stationary, and there is no correlation among hops and subcarriers.
\item Nodes can only communicate with their adjacent nodes, and cross-hop communications do not take place.
\item Half-duplex relaying is in use for all relays in all cases.
\end{itemize}

\begin{figure*}[t!]
    \centering
    \begin{subfigure}[t]{0.5\textwidth}
        \centering
        \includegraphics[width=2.8in]{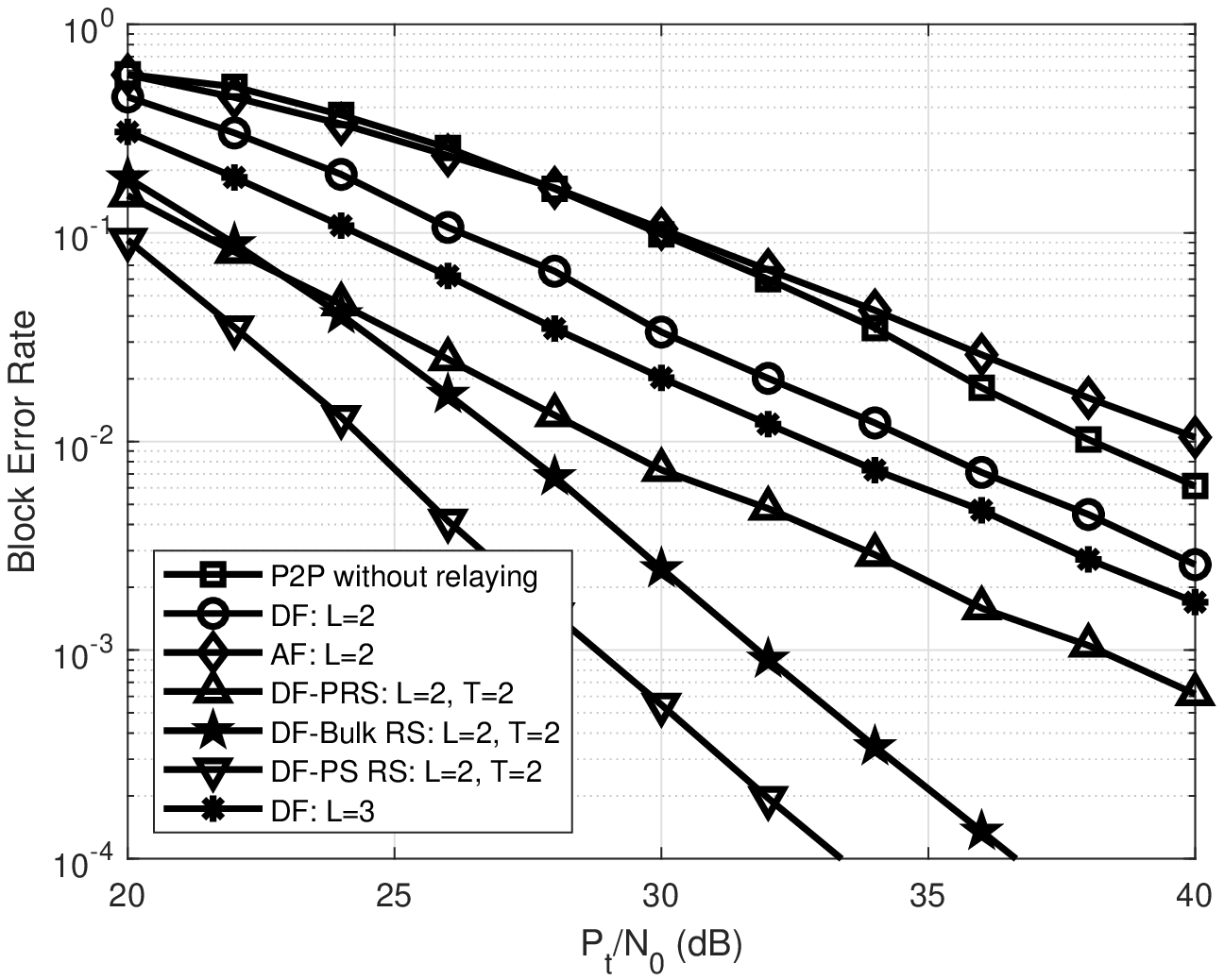}
        \caption{}
    \end{subfigure}%
~
    \begin{subfigure}[t]{0.5\textwidth}
        \centering
        \includegraphics[width=2.8in]{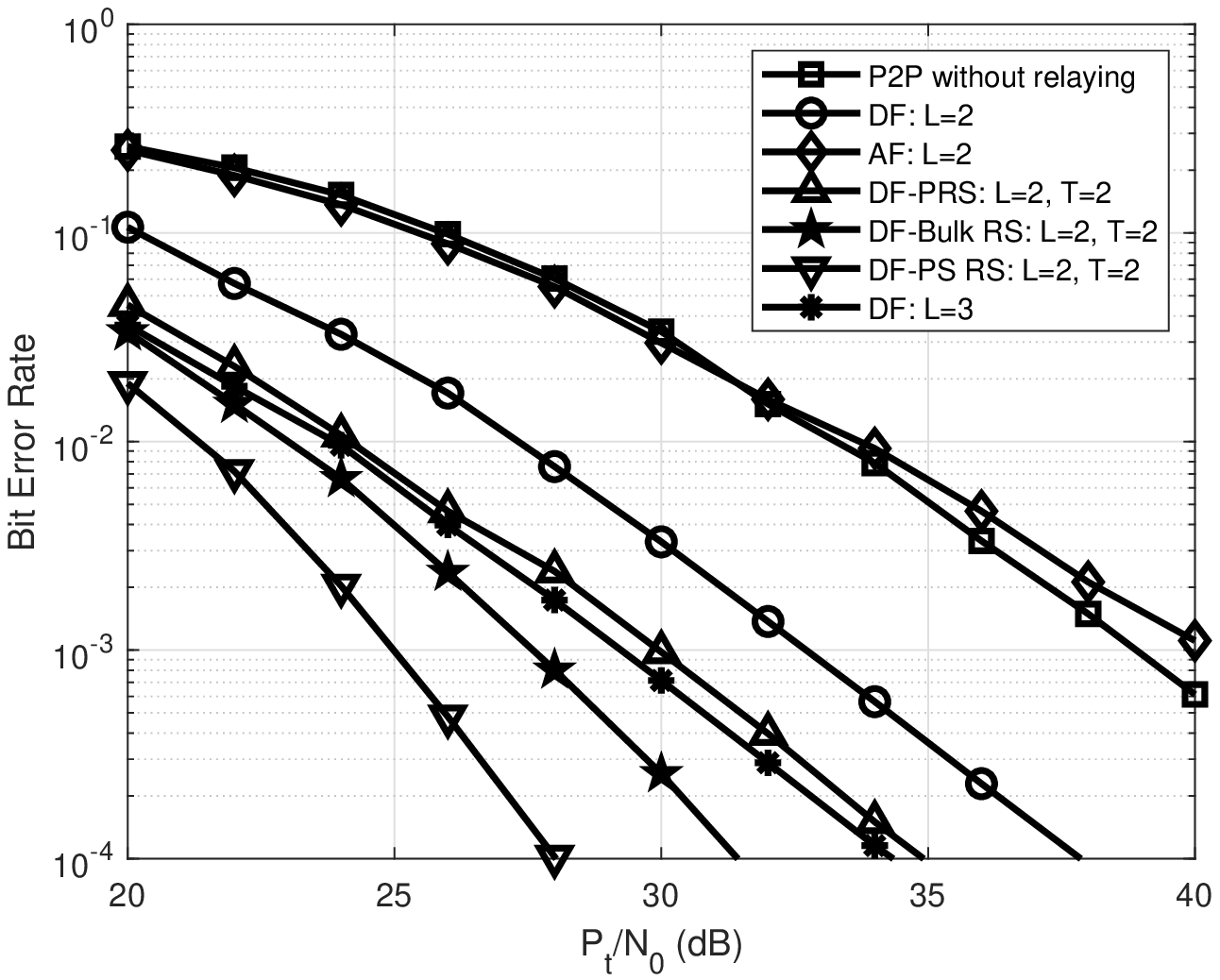}
        \caption{}
    \end{subfigure}
~
    \centering
    \begin{subfigure}[t]{0.5\textwidth}
        \centering
        \includegraphics[width=2.8in]{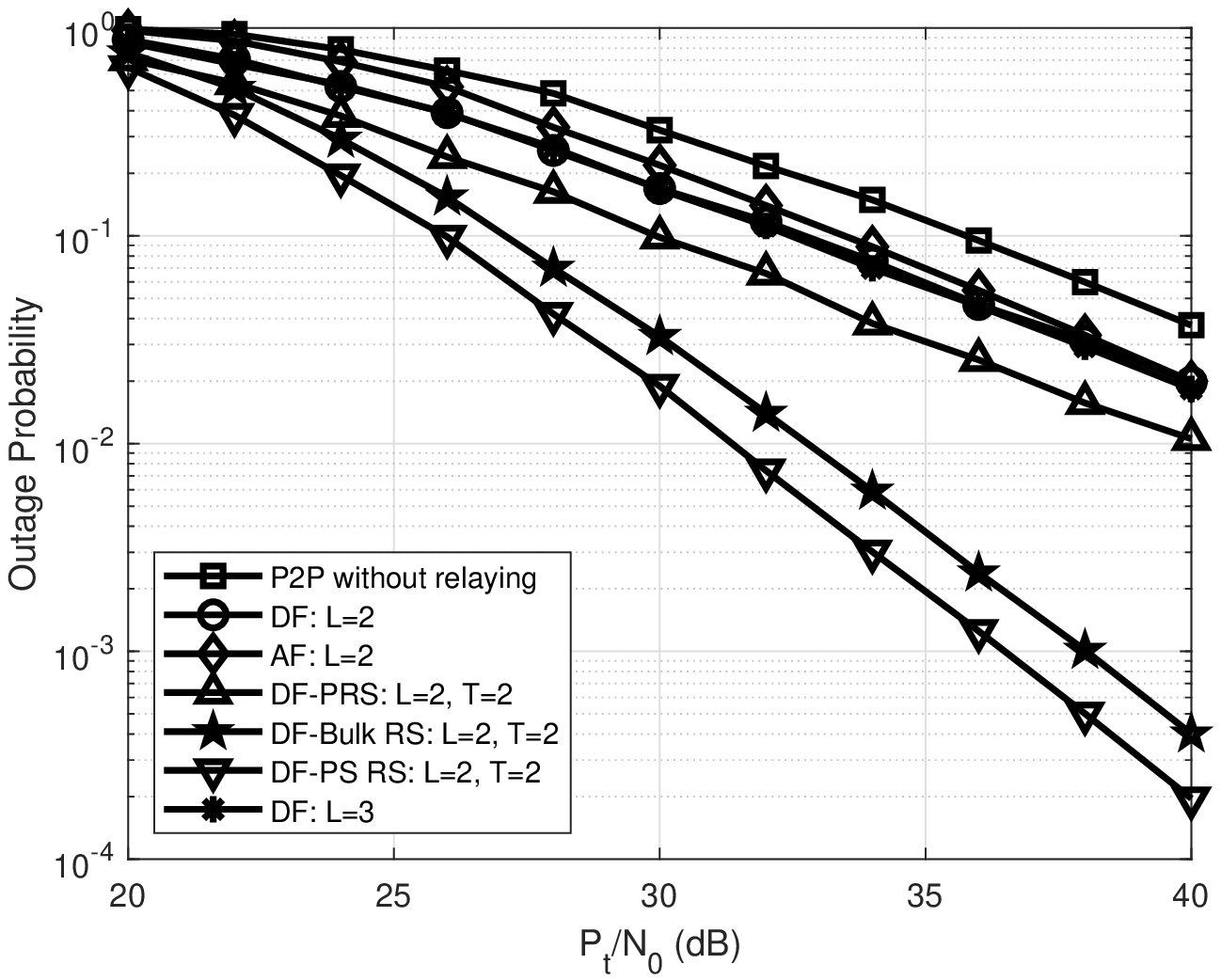}
        \caption{}
    \end{subfigure}%
~
    \begin{subfigure}[t]{0.5\textwidth}
        \centering
        \includegraphics[width=2.8in]{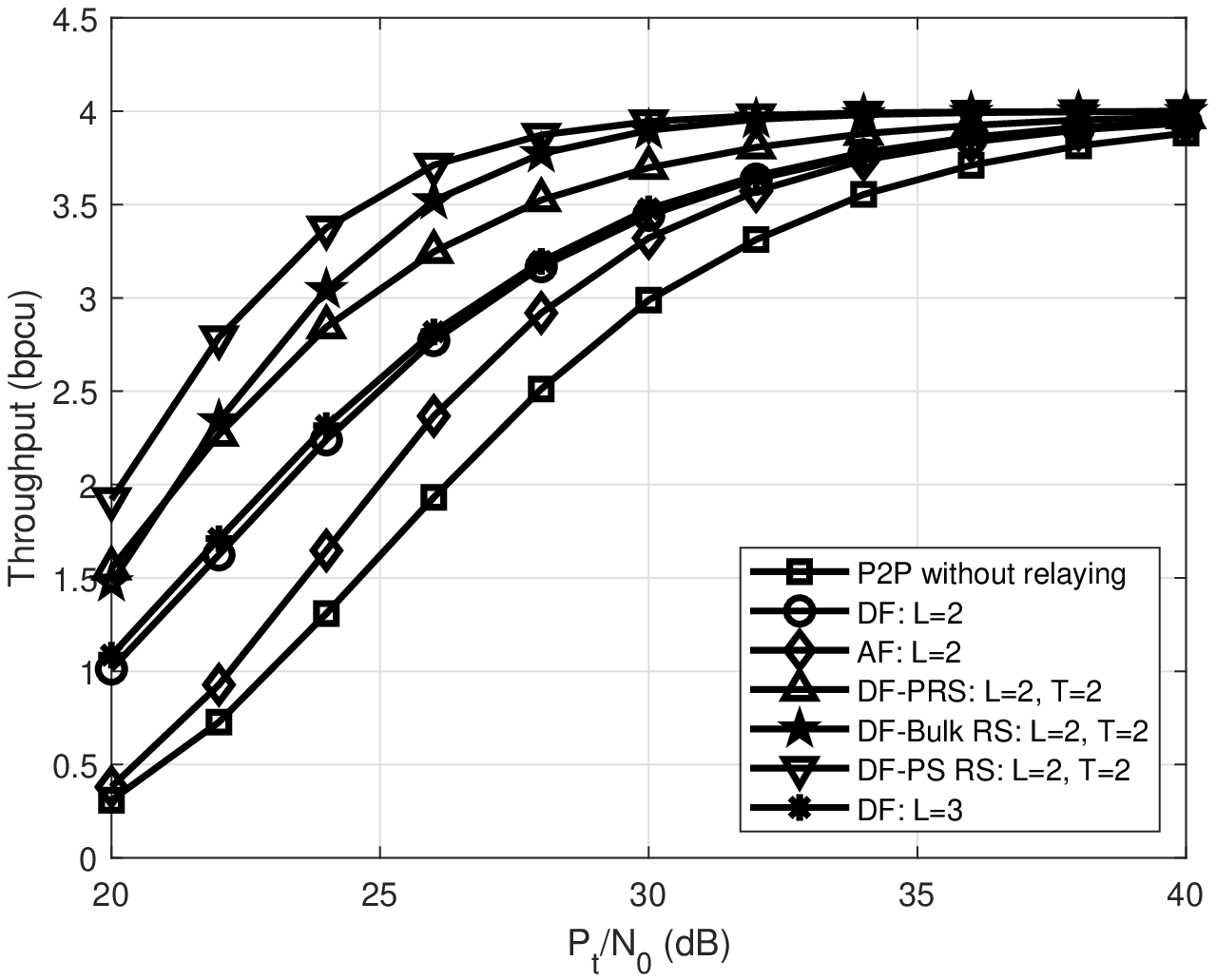}
        \caption{}
    \end{subfigure}
    \caption{Performance comparisons among point-to-point (P2P) plain OFDM-IM system and various relay assisted OFDM-IM systems with different system configurations: a) BLER; b) BER; c) OP; d) throughput.}
    \label{performance_simu}
\end{figure*}

Base on these assumptions, we set the simulation platform with a fixed network topology as follows. The source and destination are separated by $d_{SD}=10$~m without direct transmission link. Depending on the number of hops $L$, nodes are uniformly distributed over the straight line connecting the source and destination with separation $d_{SD}/L$. Transmit power $P_t$ adopted at all nodes will be uniformly distributed over $K$ active subcarriers. Binary PSK (BPSK) is used for APM over each individual active subcarrier ($M=2$); there are $N=4$ subcarriers in total, from which $K=2$ subcarriers will be activated each time for transmission and indexing purposes. Small-scale fading variation, noise power, and outage threshold are all normalized. The numerical results based on this simulation platform are generated by Monte Carlo methods and presented in Fig. \ref{performance_simu}. 

By observing the simulation results presented in Fig. \ref{performance_simu}, we can find a series of key properties of relay assisted OFDM and explore the effects of system configurations on performance. First, for most cases, cooperative relaying is beneficial to OFDM-IM, because a longer propagation distance can be split into $L$ segments and for each segment, the large-scale fading caused by path loss can be mitigated. Second, forwarding protocols are crucial and should be elaborately selected. DF relaying owns better performance than AF relaying at the cost of a higher complexity and instantaneous CSI. On the other hand, AF relaying could deteriorate the error performance of OFDM-IM, since the noise will also be amplified and could lead to a considerably distorted received OFDM block at the destination and thus a more error-prone system. Besides, when $d_{SD}$ is given, a larger number of hops $L$ will result in better performance, as more communication resources (relay nodes and power) are invested to handle the transmission. 

Meanwhile, by applying three different RS schemes, one can also have an insight into the effects of RS on the performance of relay assisted OFDM-IM. PRS scheme only considering the channels over the first hop cannot achieve a diversity gain for relay assisted OFDM-IM, but a coding gain is attainable. Because bulk and PS RS schemes view the fading in an end-to-end manner, a diversity gain equaling the number of relays for selection can therefore be harvested. Furthermore, because PS RS scheme has a higher degree of freedom by allowing selecting multiple relays, a higher coding gain is obtained than that of bulk RS scheme.

\section*{Challenges and Future Work}
As relay assisted OFDM-IM is a young-born paradigm and has attached researchers' attention just since last year, there still exist a number of challenges and interesting research topics awaiting exploration. We summarize them as follows to accelerate further related research activities.

\subsection*{Optimal Deployment of Relay Nodes}
In current literature, the deployment related issues of relay nodes, including geographical distributions and locations are neglected. As relays might not always bring benefit to OFDM-IM systems without an appropriate deployment scheme, these deployment related issues are of high importance. Poisson point process (PPP) would be utilized to model the random distribution of relay nodes over a two-dimensional plane. Mode selection between direct transmission and cooperative transmission modes should also be enabled subject to relays' physical locations and CSI. Besides, as there exists a performance trade-off in the number of hops $L$, how many hops should be involved for one complete transmission by relay assisted OFDM-IM is still an open optimization problem.

\subsection*{Realistic Channel Modeling}
In most existing papers of relay assisted OFDM-IM, due to the analytical simplicity, Rayleigh fading model is adopted in most cases and an ideal environment is taken into account, where there is not interference and correlation in all domains. To be more realistic, other fading models, for example Rician, Nakagami, and Weibull fading models shall be investigated depending on the signal propagation environment. Meanwhile, without proper measures, a variety of interference would exist and should be modeled as random variables. Even through frequency correlation can be eliminated by interleaved grouping \cite{6841601}, correlation in the spatial domain is also possible when relays are close to each other and/or move at a high speed, which will lead to sophisticated multi-hop channel modeling.

\subsection*{Synchronization and Coordination}
Synchronization and coordination among multiple relays are always indispensable in order to fully exploit the advantages of relay assisted OFDM. Both regulate the relays by when and how to collaborate so as to optimize one or several performance gains. However, in all existing papers, they are simply assumed to be perfectly arranged, which might not be the case in realistic environments. Further research activities considering both issues are worthwhile.

\subsection*{Multi-User Scenarios and Resource Allocations}
All existing papers consider a single user pair, except in \cite{8227748} both primary and secondary transmissions are analyzed. More generally, it would be interesting to extend the relay assisted OFDM-IM to multi-user scenarios. As a consequence, the resultant resource allocation problems have at least three dimensions: relay, subcarrier and power, that is, which user pairs should activate which subcarriers to transmit via which relays by how much amount of transmit power. Consequently, this extension could bring relay assisted OFDM-IM to a new realm and is therefore worth studying.

\subsection*{Relay Assisted OFDM-IM with Advanced OFDM-IM Schemes}
Apart from the original OFDM-IM scheme relying on the look-up table method and the combinatorial method proposed in \cite{6587554}, there are also a plenty of advanced and derivative OFDM-IM schemes, for example OFDM with generalized IM (OFDM-GIM), OFDM with precoded IM (OFDM-PIM), enhanced OFDM-IM (E-OFDM-IM), vector OFDM-IM (V-OFDM-IM), Differential OFDM-IM (D-OFDM-IM), multiple-mode OFDM-IM (MM-OFDM-IM), and so on (c.f. \cite{8004416} for more details). Because they are not conflict with the cooperative multi-hop architecture in essence, cooperative relaying would also be incorporated with these novel OFDM-IM schemes, so as to achieve better performance.

\subsection*{System-Level Implementation and Verification}
Although analysis and numerical results generated by simulations have proved the feasibility and superiority of relay assisted OFDM-IM in a number of application scenarios. In order to fully capture the characteristics of relay assisted OFDM-IM and testify its values in practical wireless communication systems, relevant system-level experiments on laboratory testbeds are required, but have not been done yet.

\section*{Conclusion}\label{c}
In this article, we first introduced the basics of relay assisted OFDM-IM applied over three major cooperative structures. Following the basics, we summarized the state-of-the-art achievements associated with this new paradigm in recent years via a brief literature review. By the literature review, the advantages and disadvantages of relay assisted OFDM-IM have been summarized and demonstrated by a series of comprehensive simulations. Finally, to promote relay assisted OFDM-IM and accelerate the related research activities, we outlined the existing challenges and revealed potential research directions for future work.

\bibliographystyle{IEEEtran}
\bibliography{bib}

\begin{IEEEbiographynophoto}{Shuping Dang} [M'18] (shuping.dang@kaust.edu.sa) received a B.Eng (Hons) in Electrical and Electronic Engineering from the University of Manchester (with first class honors) and a B.Eng in Electrical Engineering and Automation from Beijing Jiaotong University in 2014 via a joint `2+2' dual-degree program, and a D.Phil in Engineering Science from University of Oxford in 2018. He joined in the R\&D Center, Huanan Communication Co., Ltd. after graduating from University of Oxford, and is currently working as a Postdoctoral Fellow with the Computer, Electrical and Mathematical Science and Engineering (CEMSE) Division, King Abdullah University of Science and Technology (KAUST).
\end{IEEEbiographynophoto}
\begin{IEEEbiographynophoto}{Basem Shihada}[SM'12] (basem.shihada@kaust.edu.sa) is an associate and founding professor of computer science and electrical engineering in the Computer, Electrical and Mathematical Sciences and Engineering (CEMSE) Division at King Abdullah University of Science and Technology (KAUST). Before joining KAUST in 2009, he was a visiting faculty at the Computer Science Department in Stanford University. His current research covers a range of topics in energy and resource allocation in wired and wireless communication networks, including wireless mesh, wireless sensor, multimedia, and optical networks. He is also interested in SDNs, IoT, and cloud computing. In 2012, he was elevated to the rank of Senior Member of IEEE.
\end{IEEEbiographynophoto}
\begin{IEEEbiographynophoto}{Mohamed-Slim Alouini}[F'09] (slim.alouini@kaust.edu.sa) received the Ph.D. degree in electrical engineering from the California Institute of Technology (Caltech), Pasadena, CA, USA, in 1998. He served as a faculty member at the University of Minnesota, Minneapolis, MN, USA, then at Texas A\&M University at Qatar, Education City, Doha, Qatar, before joining King Abdullah University of Science and Technology (KAUST), Thuwal, Makkah Province, Saudi Arabia as a professor of electrical engineering in 2009. At KAUST, he leads the Communication Theory Lab and his current research interests include the modeling, design, and performance analysis of wireless communication systems.
\end{IEEEbiographynophoto}

\end{document}